\iftrue
\documentclass[aps,prl,twocolumn,
			   groupedaddress,superscriptaddress,
			   amsfonts,amssymb,amsmath,
			   citeautoscript,
			   a4paper]{revtex4-1}
\else
\documentclass[aps,pra,preprint,
			   groupedaddress,superscriptaddress,
			   amsfonts,amssymb,amsmath,
			   citeautoscript,
			   a4paper]{revtex4-1}
\fi

\usepackage[pdftex]{hyperref}
\hypersetup{colorlinks,
			linkcolor={blue!75!black!80!yellow},
			citecolor={blue!75!black!80!yellow},
			urlcolor={blue!75!black!80!yellow},
			pdfstartview=FitH}
\usepackage{siunitx}
\usepackage{graphicx}
\usepackage{mathrsfs}
\usepackage{xspace}
\usepackage{braket}
\usepackage{xr}
\usepackage{gensymb} 
\usepackage{xcite}
\usepackage{xcolor,soul}
\usepackage{stmaryrd} 
\usepackage[UKenglish]{babel}
\usepackage{placeins} 
\usepackage{bbm}
\usepackage{physics}

\usepackage{siunitx}
\DeclareSIUnit[number-unit-product=]\percent{\char`\%} 

\usepackage{txfonts}  
\usepackage{txfontsb} 

\makeatletter
\newcommand*{\addFileDependency}[1]{
  \typeout{(#1)}
  \@addtofilelist{#1}
  \IfFileExists{#1}{}{\typeout{No file #1.}}
}
\makeatother

\makeatletter
\renewcommand\@make@capt@title[2]{%
	\@ifx@empty\float@link{\@firstofone}{\expandafter\href\expandafter{\float@link}}%
	\sffamily{\textbf{#1}}\@caption@fignum@sep#2
}%

\makeatother

\newcommand{\appropto}{\mathrel{\vcenter{
			\offinterlineskip\halign{\hfil$##$\cr
				\propto\cr\noalign{\kern2pt}\sim\cr\noalign{\kern-2pt}}}}}

\newcommand{\gqu}{g_\mathrm{Q}}

\newcommand{\Tmatrix}{\mathbbm{T}}

\newcommand{\ET}{\textit{E}_{\rm{T}}}
\newcommand{\Ezero}{\textit{E}_{\rm{0}}}

\usepackage{textcomp} 
\usepackage{xifthen}
\usepackage{etoolbox}
\newboolean{togglecomments}
\newboolean{togglechanges}

\setboolean{togglecomments}{true}
\setboolean{togglechanges}{false}

\newcommand{\comment}[2]{%
    \ifbool{togglecomments}%
    {\textcolor{blue!70!black}{\small\textsf{%
    \textsuperscript{\textsc{\textsf{\MakeLowercase{#1}}}}%
    [#2]}}} 
    {}}     
\newcommand{\swap}[2]{\ifbool{togglechanges}
    {#2}  
    {\textcolor{red!70!black}{[#1]}\textrightarrow{}\textcolor{green!50!black}{[#2]}}}
\newcommand{\remove}[1]{\ifbool{togglechanges}
    {}    
    {\textcolor{red!70!black}{#1}}}
\newcommand{\inset}[1]{\ifbool{togglechanges}
    {#1}  
    {\textcolor{green!50!black}{#1}}}

\newcommand{\insetODM}[1]{\ifbool{togglechanges}
    {#1}  
    {\textcolor{red!50!green}{#1}}}
\newcommand{\insetIK}[1]{\ifbool{togglechanges}
    {#1}  
    {\textcolor{blue!40!white}{#1}}}
\newcommand{\insetZTX}[1]{\ifbool{togglechanges}
    {#1}  
    {\textcolor{cyan!60!green}{#1}}}

\newcommand{\optional}[1]{\ifbool{togglechanges}
    {}    
    {\textcolor{yellow!50!orange!80!gray}{#1}}}

\newcommand{\citeremind}[1]{%
    [\textcolor{blue!75!black!80!yellow}{
        $\blacksquare$%
	    \ifthenelse{\isempty{#1}}
	        {}
	        {\textsuperscript{\tiny\textsf{#1}}}%
	}]\xspace}

\newcommand{\eg}{e.g.\@\xspace}

\newcommand{\hkuaffil}{\footnotesize Department of Physics and HK Institute of Quantum Science and Technology,
The University of Hong Kong, Pokfulam, Hong Kong, China}
\newcommand{\HUDAaffil}{\footnotesize 
College of Mechanical and Vehicle Engineering, Hunan University, Changsha, Hunan Province, China}
\newcommand{\HUDaffil}{\footnotesize
Greater Bay Area Institute for Innovation, Hunan University, Guangzhou, Guangdong Province, China}

\begin{document}

\title{
Synthetic gain for electron-beam spectroscopy
}


\author{Yongliang~Chen}
\thanks{Y.~C. and K.~Z. contributed equally to this work.}
\affiliation{\hkuaffil}
\author{Kebo~Zeng}
\thanks{Y.~C. and K.~Z. contributed equally to this work.}
\affiliation{\hkuaffil}
\author{Zetao~Xie}
\affiliation{\hkuaffil}
\author{Yixin~Sha}
\affiliation{\hkuaffil}
\author{Zeling~Chen}
\affiliation{\hkuaffil}
\author{Xudong~Zhang}
\affiliation{\hkuaffil}
\author{Shu~Yang}
\affiliation{\hkuaffil}
\author{Shimeng Gong}
\affiliation{\HUDAaffil}\affiliation{\HUDaffil}
\author{Yiqin~Chen}
\affiliation{\HUDAaffil}\affiliation{\HUDaffil}
\author{Huigao Duan}
\affiliation{\HUDAaffil}\affiliation{\HUDaffil}
\author{Shuang~Zhang}
\email{shuzhang@hku.hk}
\affiliation{\hkuaffil}
\author{Yi~Yang}
\email{yiyg@hku.hk}
\affiliation{\hkuaffil}

\begin{abstract}
Electron-beam microscopy and spectroscopy featuring atomic-scale spatial resolution have become essential tools used daily in almost all branches of nanoscale science and technology.
As a natural supercontinuum source of light, free electrons couple with phonons, plasmons, electron-hole pairs, inter- and intra-band transitions, and inner-shell ionization.
The multiple excitations, intertwined with the intricate nature of nanostructured samples, present significant challenges in isolating specific spectral characteristics amidst complex experimental backgrounds.
Here we introduce the approach of synthetic complex frequency waves to mitigate these challenges in free-electron--light interaction. 
The complex frequency waves, created through causality-informed coherent superposition of real-frequency waves induced by free electrons, offer virtual gain to offset material losses. This amplifies and enhances spectral features, as confirmed by our electron energy loss and cathodoluminescence measurements on multi-layer membranes, suspended nanoparticles, and film-coupled nanostructures.
Strikingly, we reveal that our approach can retrieve resonance excitation completely buried underneath the zero-loss peak,  substantially enhance the quality of hyperspectral imaging, and resolve entangled multiple-photon--electron events in their quantum interaction. 
Our findings indicate the versatile utility of complex frequency waves in various electron-beam spectroscopy and their promising diagnostic capabilities in free-electron quantum optics.
\end{abstract}

\maketitle

Free-electron--light interactions~\cite{de2010optical,polman2019electron,garcia2021optical,coenen2017cathodoluminescence,talebi2018electron,rivera2020light,shiloh2022miniature,roques2023free} lay foundations for advanced electron-beam microscopy and spectroscopy~\cite{de2010optical,polman2019electron,garcia2021optical,coenen2017cathodoluminescence,hassan2017high,varkentina2022cathodoluminescence,nabben2023attosecond,bucher2024coherently,gaida2024attosecond}, tunable radiation sources spanning from X-ray, ultraviolet, Terahertz to microwave regime~\cite{massuda2018smith,roques2019towards,liu2017integrated,korbly2005observation,shentcis2020tunable,yang2023photonic,gong2023interfacial,huang2023quantum,roitman2024coherent}, integrated dielectric accelerators~\cite{sapra2020chip,shiloh2021electron,chlouba2023coherent}, free-electron lasers~\cite{pellegrini2016physics}, and quantum optics advances based on free-electron--photon entanglement~\cite{bendana2011single,kfir2019entanglements,di2019probing,di2020free,ben2021shaping,di2021modulation,dahan2021imprinting,kfir2021optical,karnieli2021coherence,konevcna2022entangling,baranes2022free,feist2022cavity,adiv2023observation,synanidis2024quantum,rasmussen2024generation}.  
Among these applications, electron energy-loss spectroscopy (EELS) and cathodoluminescence (CL) have arguably generated the most profound influence: they have evolved into daily-use methodology in physics, chemistry, and biomedicine because of their superior atomic-scale spatial resolution \cite{de2010optical,polman2019electron,garcia2021optical}.
Moreover, photon-induced near-field electron microscopy (PINEM) ~\cite{barwick2009photon,park2010photon,garcia2010multiphoton} combines ultrafast optics to electron microscopy and enables the measurement capability for free-electron--based ultrafast interaction dynamics on a sub-femtosecond time scale~\cite{feist2015quantum,priebe2017attosecond,di2019probing,vanacore2019ultrafast,synanidis2024quantum}.
In a similar vein, electron energy gain spectroscopy (EEGS) aims to enhance spectral resolution by employing pulsed illumination with tunable frequency, reaching deep into the sub-microelectronvolt energy range~\cite{auad2023muev,asenjo2013plasmon,de2008electron}.

Despite these rapid advances, the naturally weak interaction between free electrons and photons, as evident by the small magnitude of the fine structure constant, leads to measurement challenges regularly encountered in electron microscopy and spectroscopy, where spectral features could be hard to resolve or even lost due to their limitations listed below. 
First, one has to deal with a low signal-to-noise ratio (SNR)~\cite{hofer2016fundamentals,de2010optical}, especially when a high energy resolution is used in EELS. Meanwhile, CL probability suffers from low emission efficiency, resulting in a stringent SNR condition in CL spectroscopy.
Second, because free-electron excitation naturally contains all frequency components, adjacent spectral features could be partially masked by or buried in the tails of their neighbors. 
This is particularly true for the zero-loss peak notorious for overshadowing low-energy excitations (covering the terahertz, infrared, and optical regimes), a unique limitation of electron spectroscopy that one needs to battle with compared with optical spectroscopy.
Third, spontaneous events of multiple entangled photons~\cite{garciía2013multiple,feist2022cavity,adiv2023observation,arend2024electrons} are being heavily sought in free-electron quantum optics. %
Specifically, these multi-photon events obey the Poisson distribution, and thus, their probability becomes increasingly weak and calls for methods to confirm their experimental signatures.

One potential way to mitigate these limitations is to compensate for the intrinsic material losses such that spectral features become more pronounced. 
To this end, one may assemble real gain into samples to offset the loss~\cite{grgic2012fundamental,fang2009self,ramakrishna2003removal}; 
however, the addition of gain may induce instability and noise, and certain types of samples, such as those sensitive to heat, can be intrinsically incompatible with real gain.
An alternative approach is introducing complex frequency wave (CFW) with virtual gain~\cite{archambault2012superlens,tetikol2020enhancement,li2020virtual,kim2022beyond}. CFW with temporal attenuation has been shown to have the capability for compensating the loss in optical~\cite{tetikol2020enhancement,li2020virtual,kim2022beyond,kim2023loss} and acoustic systems~\cite{gu2022transient,kim2023loss}.
Recent advances have demonstrated a synthetic CFW approach featuring causality-informed integration of multiple real-frequency excitations. This synthetic approach has been demonstrated in enhancing superlens imaging~\cite{guan2023overcoming}, phonon polariton propagation~\cite{guan2024compensating}, and molecular sensing~\cite{zeng2024synthesized}. 
However, this progress all focuses on optical microscopy and spectroscopy, whereas their electron-beam counterpart remains unexplored.

\begin{figure*}[htbp]
	\centering
    \includegraphics[width=\linewidth]{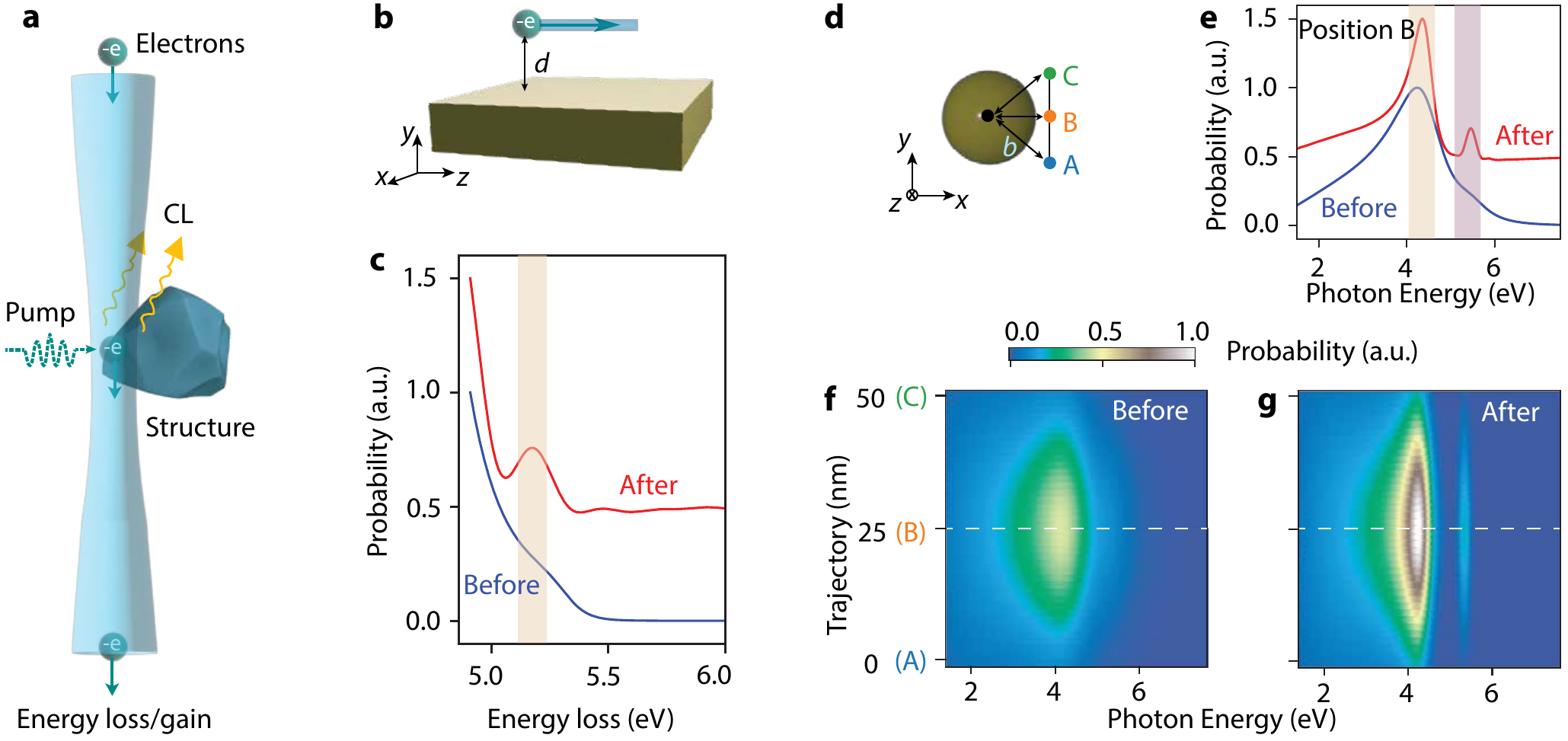}
	 \caption{%
	 	\textbf{Theoretical analysis of synthetic gain for electron spectroscopy.} 
            \textbf{a.} General schematic of electron energy gain and loss spectroscopy and cathodoluminescence: an electron beam interacts with an arbitrary structure in either aloof or penetrating configurations with optional external laser pumping. 
            \textbf{b-c.} Analytical scenario (b) of 200-keV electrons moving along the $z$ direction in vacuum above a Drude-metal half-space with a separation of $d=\SI{100}{\nm}$ and the associated EELS spectra (c; before and after CFW processing are denoted by blue and red color, respectively). 
            The surface plasmon loss near \SI{5.2}{\eV} (see brown shading), previously hidden within the tail of the zero-loss peak, gets retrieved after CFW processing.
            \textbf{d-e.} Analytical scenario (d) of 10-keV electrons moving along the $z$ direction in vacuum near a Drude-metal nanosphere of diameter \SI{50}{\nm} and the associated CL spectra (e; before and after CFW processing are denoted by blue and red color, respectively) at Position B with an impact parameter $b=\SI{55}{\nm}$. 
            \textbf{f-g.} CL hyperspectral images before (left) and after (right) CFW processing when the electron beam scans from Positions A ($b=\SI{74}{\nm}$) to B ($b=\SI{55}{\nm}$) and C ($b=\SI{74}{\nm}$) in d. 
            The horizontal cuts (dashed white line) in f and g correspond to e.
        }
	\label{fig_main: schematic}
\end{figure*}

%
This work introduces the CFW approach to free-electron--light interaction for enhancing electron-beam microscopy and spectroscopy with synthetic gain. 
After theoretically showcasing its effectiveness with canonical geometries, we demonstrate the general utility of this approach using three sets of measurements, including EELS measurements of a dielectric-metal-dielectric (DMD) multi-layer membrane, EELS measurement of single silver nanoparticles, and CL spectroscopy of film-coupled plasmonic nanospheres.
We further apply our approach to electron energy gain spectroscopy and quantum-entangled free-electron multi-photon events.
These efforts jointly show that the CFW approach enables the retrieval of weak hidden spectral features and the enhancement of the existing features, thereby substantially improving the capability for mode characterization, hyperspectral imaging, and quantum event detection in electron-beam microscopy and spectroscopy.
%

\begin{figure*}[htbp]
	\centering
    \includegraphics[width=1\linewidth]{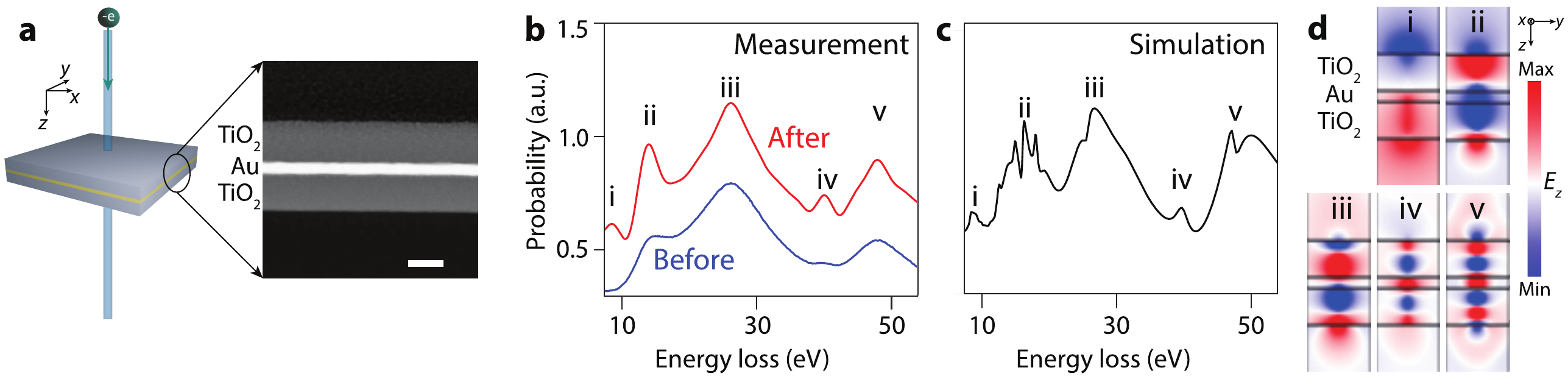}
	 \caption{%
	 	\textbf{Enhanced electron energy loss spectroscopy of a suspended dielectric-metal-dielectric (DMD) membrane.}             %
            \textbf{a.} Schematic (left) and cross-sectional transmission electron microscope (TEM) image (right) of a 200-keV electron beam penetrating through a DMD membrane consisting of a gold layer (\SI{6}{nm} thickness) sandwiched by TiO$_2$ (\SI{30}{nm} thickness for both superstrates and substrates). Colorbar: \SI{20}{\nm}.
            \textbf{b.} Measured spectra before (blue) and after (red) CFW processing. Peaks i and iv, particularly unnoticeable in the raw spectrum, become evident after processing. 
            \textbf{c-d.} Simulated spectrum (c) and field profiles (d) of the five spectral peaks.
	    }
	\label{fig_main:DMD}
\end{figure*}

In EELS and CL, electron beams generate evanescent incident fields spanning all frequencies, interact with and get scattered by specimens, and lead to electron energy loss and radiation known as cathodoluminescence, as shown in Fig.~\ref{fig_main: schematic}a.
Broadly speaking, it has been established that a passive scatterer in a vacuum is causal, meaning that it is analytic in the upper-half plane and satisfies the Kramers–Kronig (KK) relations~\cite{srivastava2021causality}.
The free-electron manifestation of this principle can be mostly conveniently seen in small point particles, whose energy loss is simply proportional to the imaginary part of their polarizability~\cite{polman2019electron}.
For general situations, one could leverage a $\Tmatrix$-matrix--based oscillator representation framework~\cite{zhang2023all,xie2024maximalquantuminteractionfree} (where the $\Tmatrix$ operator relates the polarization fields with incident fields~\cite{carminati2021principles,chao2022physical}) and show that the energy loss is proportional to the imaginary part of the $\Tmatrix$ matrix (see Sec. S5), which is causal and passive simultaneously.

Based on this foundation, to augment the features in electron microscopy and spectroscopy, one potential mathematical method is to replace real frequency with a suitable complex frequency $\tilde\omega = \omega-\mathrm{i}\tau /2$ for loss compensation. 
This is equivalent to offset the imaginary part of material permittivity, \eg, transforming the permittivity of a Lorenz model $\varepsilon = 1 - \omega_p^2/(\omega^2+ \mathrm{i}\omega\gamma - \omega_0^2)$ into a purely real value, $\varepsilon(\tilde\omega) = 1 - \omega_p^2/(\omega^2+ \gamma^2/4 - \omega_0^2)$ by the substitution of $\omega$ with $\omega-\mathrm{i}\gamma /2$.
However, in this treatment, energy would diverge as time tends to negative infinity, indicating that the theoretically ideal CFW is unphysical.
Practically, we can express a time-truncated CFW with the expression of $\ET = \Ezero\mathrm{e}^{-\mathrm{i}\tilde\omega t}\Theta(t)$, where $\Ezero$ is the original signal in complex form (\eg electric fields), $\ET$ is the time-truncated signal, and $\Theta(t)$ is the Heaviside step function with $\Theta(t)=0$ at $t<0$ and $\Theta(t)=1$ otherwise.  
By leveraging the causality property of the $\Tmatrix$-matrix, we can recover a loss probability with complex values $\tilde\Gamma\left(\omega\right)$ based on the Kramers-Kronig relations, $\Re{\tilde\Gamma\left(\omega\right)} = \frac{1}{\pi} \mathcal{P} \int_{-\infty}^{\infty} \frac{\Im{\tilde\Gamma\left(\omega\right)}}{\omega' - \omega} d\omega'$, 
where $\mathcal{P}$ denotes the Cauchy principal value, and $\Im{\tilde\Gamma\left(\omega\right)}\equiv\Gamma\left(\omega\right)$
represents the real-frequency probability measured from experiments. 
Using Fourier transformation, a linear combination of real-frequency probability from experimental acquisition can be used to synthesize the complex-valued loss probability $\tilde\Gamma(\tilde\omega)$ at complex frequencies $\tilde\omega$:
\begin{align}
\tilde\Gamma(\tilde\omega)
\approx\sum_{n}\tilde\Gamma(\omega_n)\mathrm{e}^{-\mathrm{i}\omega_n t + \mathrm{i}\tilde\omega t}\Delta\omega/\left[(2\pi \mathrm{i}\left(\tilde\omega - \omega_n\right)\right]
\label{eq:T3ext}
\end{align}
and the real-valued probability associated with the complex frequency is given by $\Im{\tilde\Gamma\left(\tilde\omega\right)}$.

Next, we theoretically demonstrate the synthetic CFW approach using two analytical scenarios as shown in Fig.~\ref{fig_main: schematic}b-g.
The first example is an electron passing above a metal half-space at a distance of $d$ in vacuum, as shown in Fig.~\ref{fig_main: schematic}b, which allows for analytical treatment~\cite{lucas1971fast,echenique1975absorption,otto1967theory,garcia1985retardation,forstmann1991energy} (see Sec. S6A). 
When the separation is large at $d=\SI{100}{\nm}$, the spectral signature at \SI{5.2}{\eV} associated with surface plasmon launching, gets buried in the tail of the zero-loss peak, as seen in the blue curve before CFW processing in Fig.~\ref{fig_main: schematic}c. 
Aiming at restoring the SP peak, we apply the CFW processing based on Eq.~\eqref{eq:T3ext}. 
The post-CFW spectrum is shown by the red curve in Fig.~\ref{fig_main: schematic}c, where the previously buried surface plasmon mode can be seen clearly.
It is noted that the red curve in Fig.~\ref{fig_main: schematic}c is shifted up by 0.5 relative to zero probability for better visualization (same treatment applied to all comparisons in other figures containing multiple spectra unless specified otherwise).

The second analytical example is an electron passing near a nanoparticle with an impact parameter $b$, as shown in Fig.~\ref{fig_main: schematic}d.
The CL spectra of this interaction permit analytical treatment~\cite{de1998relativistic,de2010optical} which describes a summation of various Mie scattering channels labelled by integers $l$ (see Sec. S6B) .
When the material loss of the nanoparticle is large, the raw spectrum (blue curve in Fig.~\ref{fig_main: schematic}e; for electrons at position B in Fig.~\ref{fig_main: schematic}d) displays a single-peak lineshape corresponding to the dipolar ($l=1$) mode. 
Nevertheless, both the dipolar and quadrupolar ($l=2$) modes can both be clearly seen after CFW processing (red in Fig.~\ref{fig_main: schematic}e). 
To further demonstrate the effectiveness of the CFW method, Fig.~\ref{fig_main: schematic}f displays a theoretical one-dimensional hyperspectral mode mapping imaging from Position A to C. 
The excitation of the dipolar mode predominates, whereas the quadrupolar mode can hardly be recognized. 
By applying CFW processing into the whole hyperspectral image, the previously hidden quadrupolar mode becomes vividly seen, leading to the occurrence of clear two-mode trajectories in the post-CFW mode mapping (Fig.~\ref{fig_main: schematic}g).

\begin{figure}[!htbp]
	\centering
	\includegraphics[width=1\linewidth]{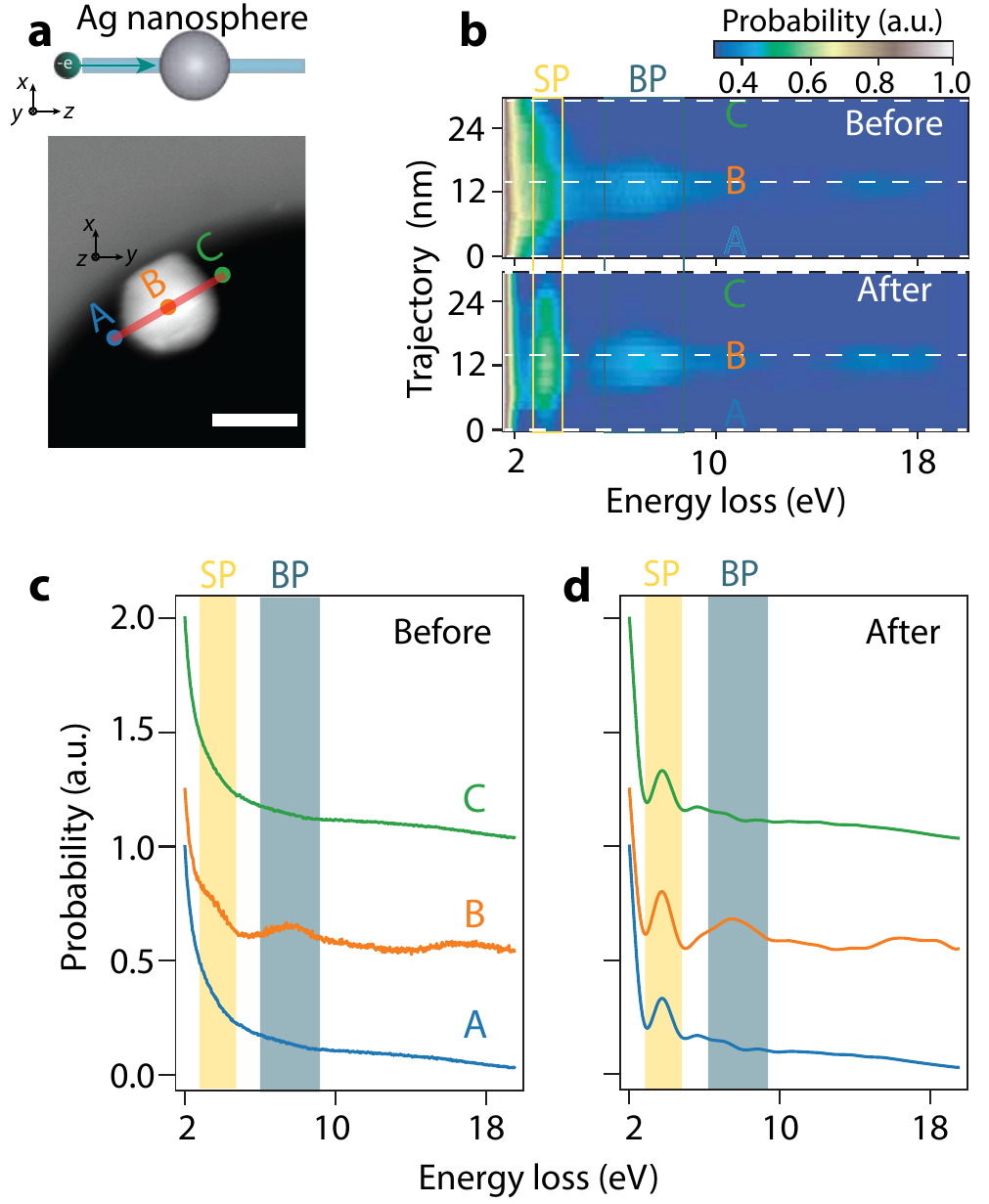}
	 \caption{%
	 	\textbf{Retrieval of localized surface plasmon resonances buried under zero-loss peak.}
	 	\textbf{a.} Schematic (top) of 200-keV electrons scanning across a single silver nanosphere of \SI{20}{\nm} diameter and the corresponding TEM image (bottom) showing the scanning line A-B-C with a step size of \SI{1}{nm} for mode mapping. Scale bar: \SI{20}{\nm}. 
            \textbf{b.} One-dimensional hyperspectral imaging with mode mapping before (top) and after (bottom) CFW processing along scanning line A-B-C in a. 
            \textbf{c-d.} EELS spectra before (c) and after (d) CFW processing acquired at Positions A, B. The colored shadings denote localized surface plasmon (SP) mode and bulk plasmon (BP) mode, respectively.
	 	}
	\label{fig_main: Ag}
\end{figure}

\begin{figure*}[htbp]
	\centering
    \includegraphics[width=1\linewidth]{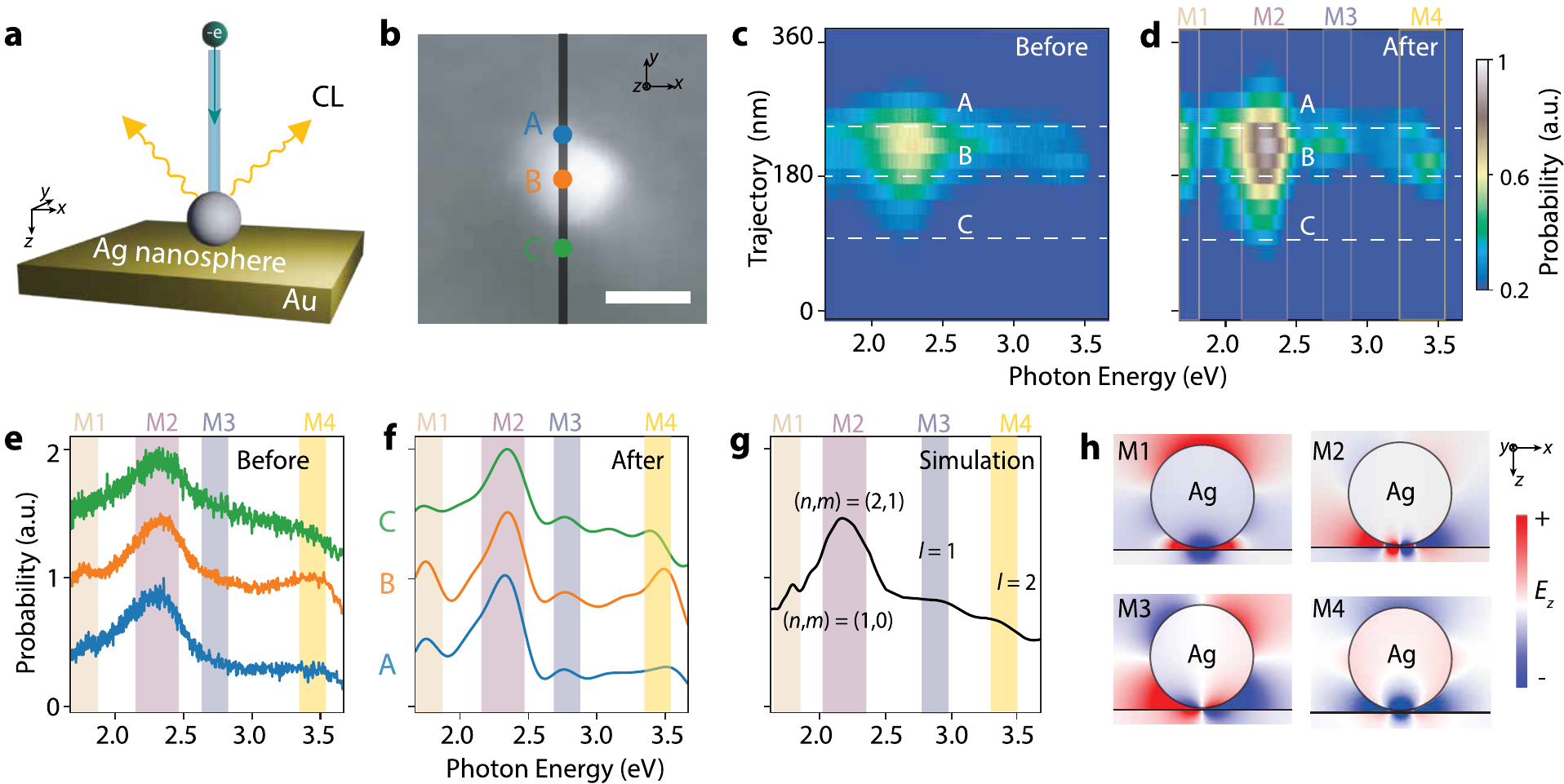}
	 \caption{%
	 	\textbf{Improved cathodoluminescence spectroscopy of a film-coupled nanosphere.
            }
            \textbf{a.} Schematic of 10-keV electrons scanning across a silver nanosphere on a gold substrate.
	 	\textbf{b.} SEM image displaying a naonsphere of \SI{100}{\nm} diameter and the A-B-C black scanning line with a \SI{15}{\nm} step size. Scale bar: \SI{100}{\nm}.
            \textbf{c-d.} One-dimensional hyperspectral imaging before (c) and after (d) CFW processing. The scanning trajectory is along the red line in b. Four modes labeled as M1 to M4 can be resolved after processing.
        \textbf{e-f.} Line cuts of c and d at Positions A, B and C.
            \textbf{g.} Simulated spectra and resonant mode identification.
            \textbf{h.} Eigenstate field profiles of the four modes. 
            }
	\label{fig_main: Ag-gold}
\end{figure*}

%
Motivated by such enticing performances as predicted in theory, we perform three measurements to showcase the capability of the CFW technique experimentally. These experiments are the energy loss measurement of a multilayer thin film and a suspended nanosphere, and the CL measurement of a film-coupled nanosphere.

We first conduct a bulk loss measurement of a dielectric-metal-dielectric (DMD) multilayer structure (Fig.~\ref{fig_main:DMD}a). 
The DMD structure consists of a \SI{6}{nm} thick Au layer sandwiched between two \SI{20}{\nm} TiO$_2$ thick layers, as shown in the cross-sectional TEM image (right of Fig.~\ref{fig_main:DMD}a). 
The originally measured bulk loss spectrum (blue curve in Fig.~\ref{fig_main:DMD}b) of 200-keV electrons displays three peaks at around \SI{12}{\eV}, \SI{25}{\eV} and \SI{48}{\eV}, respectively. 
Extra spectral features emerge in the post-CFW EELS spectrum (red curve in Fig.~\ref{fig_main:DMD}b): the original three peaks become more pronounced, and more importantly, two new peaks at around \SI{6}{\eV} and \SI{41}{\eV} appear. 
Taken together, these five peaks are labeled as i to v according to their ascending frequencies.
Furthermore, We conduct numerical simulations to calculate the energy loss probability (Fig.~\ref{fig_main:DMD}c) and induced field profiles (Fig.~\ref{fig_main:DMD}d) at those peaks.
Peak i primarily arises from the transitions from the peak of the TiO$_2$ valence band to the $T_{2g}$ and $E_g$ levels of the titanium $d$ orbitals~\cite{launay2004evidence}, with a potential contribution from the bulk plasmon excitation in gold~\cite{motornyi2020electron}.
Besides, Peaks ii and iii are mostly associated with the electron-hole pair generation in TiO$_2$~\cite{vast2002local,mo1995electronic}.
Meanwhile, Peaks iv and v stem from the titanium semicore transitions~\cite{vast2002local,wang2009crystal}.
The augmented spectral features after CFW processing are in good agreement with the numerical simulations (See Sec. S4) in Fig.~\ref{fig_main:DMD}c.
The next experiment focuses on suspended silver nanoparticles. 
Single silver nanoparticles of diameter around \SI{20}{\nm} are suspended within holes of a transmission electron microscopy (TEM) grid (Fig.~\ref{fig_main: Ag}a). 
A 200-keV electron beam scans across a single nanoparticle, creating a \SI{30}{nm} scanning line with \SI{1}{nm} steps for mode mapping. 
The hyperspectral images along the scanning trajectory from A to C (see red line in Fig.~\ref{fig_main: Ag}a) are presented in Fig.~\ref{fig_main: Ag}b (top: before CFW; bottom: after CFW). 
The EELS intensity of the peaks varies along the scanned line in accordance with the coupling efficiency at which surface and bulk plasmons are launched under different impact parameters. 
The trajectories of the surface and bulk plasmon peaks become more pronounced after the CFW processing.
To see this more clearly, Fig.~\ref{fig_main: Ag}c presents the raw EELS spectra measured at three representative positions (aloof positions A and C, and penetration position B) along the beam scanning line.
At Position B, weak features at around \SI{3.6}{\eV} and \SI{7.5}{\eV} can be seen in the EELS spectrum without CFW, corresponding to the dipolar Mie surface plasmon resonance and bulk plasmons, respectively.
Whereas at Positions A and C, spectral features cannot be recognized except for the zero-loss peak because the coupling between the electron and the nanoparticle gets weaker at larger separations.
Strikingly, after the CFW processing (Fig.~\ref{fig_main: Ag}d), the dipolar Mie resonance peak emerges in the spectra of both Position A and C, where, in the meantime, the bulk plasmon peak remains absent because the interaction is in the aloof configuration. For Position B, the weak features of the surface and bulk plasmons are both substantially enhanced.
In addition to EELS, in Sec. S7 we apply our approach to EEGS and exemplified its capability in resolving and characterizing closely neighbored resonances.

Next, we move to cathodoluminescence, another important type of spectroscopy that combines high spatial and spectral resolution.
Fig.~\ref{fig_main: Ag-gold}a illustrates the experimental structure---a film-coupled nanoantenna geometry consisting of single silver nanoparticles of $\sim$\SI{100}{\nm} diameter on a \SI{100}{nm} thick gold film, interacting with a \SI{10}{keV} electron beam. 
Again, we perform a hyperspectral CL imaging measurement along a scanning line across the nanoparticle (see black line in Fig.~\ref{fig_main: Ag-gold}b). 
In the raw hyperspectral image as measured (Fig.~\ref{fig_main: Ag-gold}c), the trajectory of only a single mode can be well recognized at \SI{2.3}{\eV}, together with faint features of another mode around \SI{3.4}{\eV}. Nevertheless, after CFW processing, four modes appear in total, and they are labeled as M1 to M4 with ascending frequency as shown in Fig.~\ref{fig_main: Ag-gold}d. 
The spectra at representative positions A, B, and C before and after CFW processing are shown in Fig.~\ref{fig_main: Ag-gold}e and f, respectively.
To elucidate the origin of each mode, we perform numerical simulations, and the resulting spectrum is shown in Fig.~\ref{fig_main: Ag-gold}g, which corresponds well with the experimental data in Fig.~\ref{fig_main: Ag-gold}f. 
We further elucidate the origin of the peaks using eigenmode calculations in Fig.~\ref{fig_main: Ag-gold}h. Because of the sphere-film--coupled geometry, the gap plasmon modes and the Mie modes get hybridized. 
In the low-frequency region, M1 and M2 predominately display gap-plasmon features and are labeled by their quantized in-plane wavevectors $(n,m)=(1,0)$ and $(2,1)$ along the radial and azimuthal direction, respectively~\cite{yang2017low}. 
As the frequency increases, M3 and M4 primarily exhibit Mie characteristics and can be identified by the integer labels $(l,m)$ of spherical harmonics. Nevertheless, CL in this region becomes highly multi-channel in $m$. Therefore, in Fig.~\ref{fig_main: Ag-gold}, we display the eigen-fields of $(l,m)=(1,1)$ and $(2,0)$ for M3 and M4, respectively. These modes are the primary contributors to these two peaks according to the Mie CL calculation~\cite{de2010optical,de1998relativistic}.
%

\begin{figure}[htbp]
	\centering
	\includegraphics[width=1\linewidth]{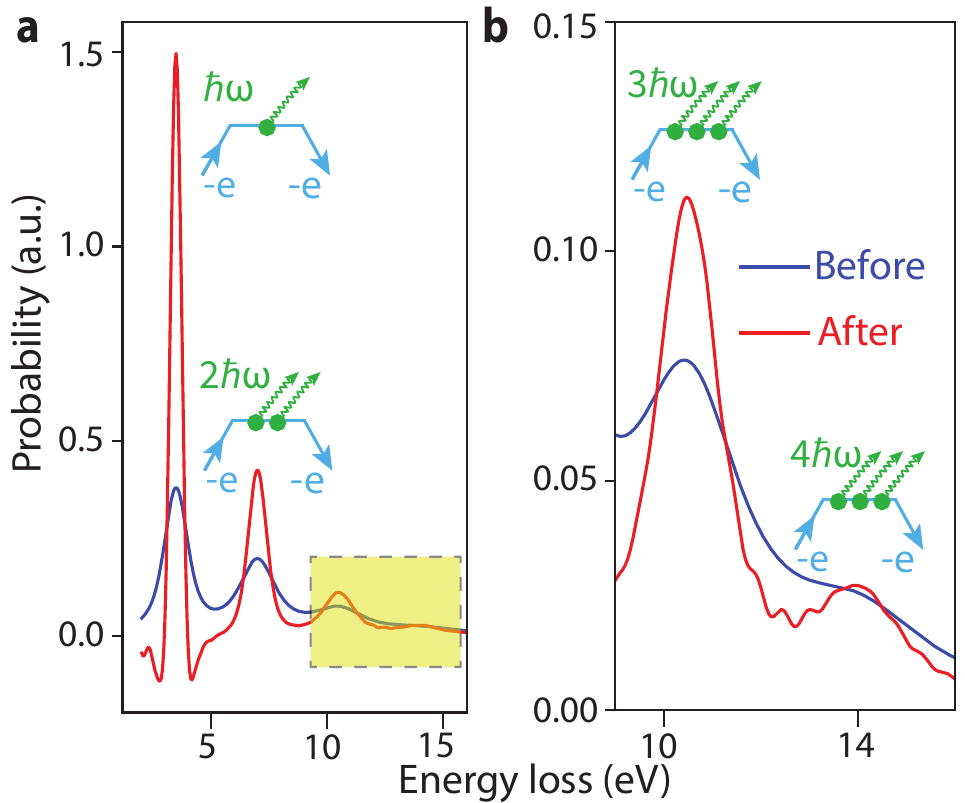}
	 \caption{%
	 	\textbf{Resolving spontaneous multiple photon events in free-electron--photon quantum interaction.
   }
            \textbf{a.} Analytically constructed electron energy loss spectrum featuring equally-spaced multiple photon loss peaks (indicated by the inset Feynman diagrams) before (blue) and after (red) CFW processing. The multi-photon probability obeys Poisson statistics in the raw spectrum.
            \textbf{b.} Zoom-in of the dashed box (shaded yellow in a) where the four-photon peak becomes clearly resolvable after processing. 
	     }
	\label{fig_main: quantum}
\end{figure}

Lastly, into the quantum interaction regime, we theoretically demonstrate that CFW can also be used for resolving experimental signatures of spontaneous free-electron multi-photon processes.
This quantum interaction can be described by a model of Poisson statistics describing the spectral density of quantized photons that are equally separated by a single photon energy in the EELS spectrum~\cite{adiv2023observation,feist2022cavity} (see Sec. S8). 
Based on this model, we construct an EELS spectrum with a large quantum coupling strength $\gqu=1$ such that multi-photon processes can spontaneously happen; the associated energy loss spectrum is shown by the blue curve of Fig.~\ref{fig_main: quantum}a. %
Despite the one-, two-, and three-photon peaks being evident, the four-photon peak (and even higher-order peaks) is poorly resolved even without noise.
In practical EELS experiments, such a weak feature will be further contaminated by the addition of experimental noise.
Nonetheless, after CFW processing, the four-photon peak becomes clearly recognizable. At the same time, all other lower-order photon peaks get enhanced (see red curves in Fig.~\ref{fig_main: quantum}a and the zoom-in image in Fig.~\ref{fig_main: quantum}b). 
This indicates that the CFW method could be particularly useful for restoring experimental signatures in free-electron quantum optics.

In conclusion, we have introduced an approach of complex frequency waves to provide virtual gain for electron microscopy and spectroscopy. 
In fact, our approach shares a similar rationale with electron energy gain spectroscopy~\cite{de2008electron}, yet differs in that EEGS relies on real gain from external laser pumping, whereas our CFW approach offers virtual gain through causality.
We have showcased its capabilities in restoring spectroscopic features and improving hyperspectral imaging across various scenarios, including multi-layer membranes, suspended and film-coupled nanoparticles, and spontaneous multi-photon processes.
Aside from the optical excitations mostly focused on in this work, our approach can be generally applied to a broad range of excitations in electron-beam spectroscopy, encompassing a spectrum ranging from low-energy phonons to high-energy ionization. 
In addition to the EELS and CL systems studied here, our CFW approach extends to optically pumped electron-beam systems like EEGS and PINEM.
Going beyond electron-beam systems, this approach may be applicable to a large family of electron spectroscopy for condensed matters, such as X-ray photoelectron spectroscopy, angle-resolved photoemission spectroscopy, scanning tunneling spectroscopy, and so forth.
The virtual gain provided by the complex frequency waves opens an avenue toward resolving subtle spectroscopic features in electron-beam microscopy and free-electron quantum optics.

\emph{Acknowledgments.}
We thank Germaine Arend, Frankie Y.~F.~Chan, Anthony H.~W, Choi, Mathieu~Kociak, Zemeng~Lin, and Bengy T.~T.~Wong for experimental help and stimulating discussions.




%

\end{document}


\title{SUPPLEMENTARY INFORMATION\\
Synthetic gain for electron-beam spectroscopy
}

\author{Yongliang~Chen}
\thanks{Y.~C. and K.~Z. contributed equally to this work.}
\affiliation{\hkuaffil}
\author{Kebo~Zeng}
\thanks{Y.~C. and K.~Z. contributed equally to this work.}
\affiliation{\hkuaffil}
\author{Zetao~Xie}
\affiliation{\hkuaffil}
\author{Yixin~Sha}
\affiliation{\hkuaffil}
\author{Zeling~Chen}
\affiliation{\hkuaffil}
\author{Xudong~Zhang}
\affiliation{\hkuaffil}
\author{Shu~Yang}
\affiliation{\hkuaffil}
\author{Shimeng Gong}
\affiliation{\HUDAaffil}\affiliation{\HUDaffil}
\author{Yiqin~Chen}
\affiliation{\HUDAaffil}\affiliation{\HUDaffil}
\author{Huigao Duan}
\affiliation{\HUDAaffil}\affiliation{\HUDaffil}
\author{Shuang~Zhang}
\email{shuzhang@hku.hk}
\affiliation{\hkuaffil}
\author{Yi~Yang}
\email{yiyg@hku.hk}
\affiliation{\hkuaffil}

\maketitle

\setlength{\parindent}{0em}
\setlength{\parskip}{.5em}

\noindent{\textbf{\textsf{CONTENTS}}}\\ 
\twocolumngrid
\begingroup 
    \let\bfseries\relax 
    \deactivateaddvspace 
    \deactivatetocsubsections 
    \tableofcontents
\endgroup
\onecolumngrid

\hyphenation{eigen-index}

\section{Sample fabrication and charcterization}
\label{sec:fab}

\subsection{Multilayer membrane}

The TiO$_2$-Au-TiO$_2$ multilayer membrane was prepared by layer-by-layer deposition onto a TEM grid (Ted Pella), which is used for one electron energy loss measurement. 
We applied the ion-beam assisted deposition technique for the preparation of the bottom and top TiO$_2$ layers. 
A quartz microbalance was used to monitor the deposition thickness. The middle ultrathin gold layer was fabricated by dual ion-beam sputtering deposition. 
To obtain a long-scale uniformity of the membrane, we adopted an ion-beam thinning-back process~\cite{ma2024pushing}, where a 10-nm thick gold layer was firstly deposited on the bottom TiO$_2$ layer by Ar$^+$ ion-beam sputtering using the main ion source and was, subsequently, thinned back to a desired thickness by ion beam polishing using the auxiliary ion source.

The thickness of the multilayer membrane was characterized by cross-section transmission electron microscopy (FEI Talos F200X) under electron energy of \SI{200}{\keV}. 
The cross-sectional sample was prepared by a focused ion beam facility (ZEISS Gemini Crossbeam 350). 
First, we aligned the cross-point of dual beams to the area of interest.
To maintain the fine multilayer structure, a platinum-carbon composite layer was coated for protection by electron-beam-induced deposition. 
Afterward, the sample was milled into a thin lamella with the desired thickness and carefully transferred onto a copper grid. 
The lamella was then attached to an Omniprobe using platinum deposition and detached by milling underneath. Finally, the lamella was further thinned using a low ion beam current to achieve electron transparency and minimize damage.

\subsection{Suspended and film-coupled nanospheres}
The single silver nanoparticle suspended in the hole of TEM grid (Ted Pella) is fabricated for the loss measurement in Fig. 3.  
20 uL solution of silver nanoparticles containing 20-30 $nm$ diameter (Biotyscience) was dropped onto the TEM grids with arrays of holes. 
Once the sample was naturally dried, the sample was stored in the TEM chamber for further measurement. 

The film-coupled nanosphere consisting of single NPs onto the gold film was fabricated for cathodoluminescence (CL) measurement as shown in Fig. 4. 
We applied the sputter-coating method for 100 nm gold layer onto one silicon substrate. 
20 uL solution containing silver nanoparticles with 90-100 nm diameter (Biotyscience) was dropped onto the resulting 100 nm gold film. 
Once the sample was naturally dried, the sample was stored in the CL-SEM chamber for further measurement.

The size of the suspended nanosphere was characterized by the high-angle angular dark field (HAADF) imaging under the electron energy of \SI{200}{keV} in the TEM. 
The size of the film-coupled nanosphere was characterized by the secondary electron (SE) imaging under the electron energy of \SI{10}{\eV} with the aperture size of \SI{100}{\micro\meter}.

\section{Measurement}
\label{sec:measure}

Electron energy loss measurements were conducted by a Talos F200X G2 Transmission Electron Microscope (TEM, ThermoFisher) equipped with an ultra-high--brightness Cold Field Emission Gun, which allows for a spatial resolution of $\leq\SI{0.1}{nm}$ and EELS energy resolution of $\leq\SI{0.3}{\eV}$. 
The one-dimensional hyperspectral imaging was acquired in a scanning mode with a step size of \SI{1}{\nm}, a spectral acquisition time of \SI{0.2}{ms}, and electron energy of \SI{200}{\keV}.

Cathdoluminescence (CL) measurements were performed by a CL-SEM (Attolight Allalin) equipped with a Schottky field emission gun and CL collection optics connected with an Andor spectrometer. The CL signals were directed by an achromatic reflective objection with a numerical aperture of 0.71 to a high-speed UV-Visible CCD camera with a wavelength from \SI{200}{\nm} to \SI{1000}{\nm}. 
The solid angle of the CL signal collected was from $\ang{20}$ to $\ang{45}$. 
The hyperspectral imaging was acquired with a step size of \SI{15}{\nm}, a spectral acquisition time of \SI{0.5}{\second}, and an electron energy of \SI{10}{\keV}.

\section{Complex frequency waves processing}
\label{sec:cfw}

%
\begin{figure*}[!htbp]
	\centering
    \includegraphics[width=1\linewidth]{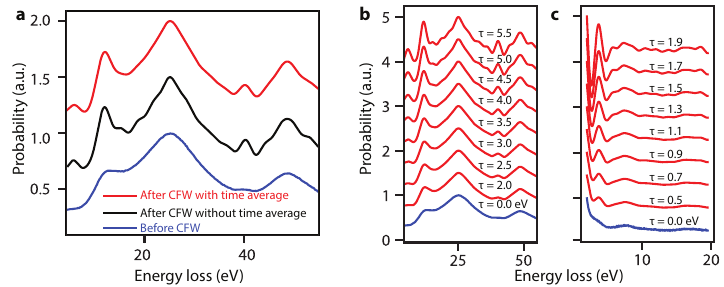}
	 \caption{%
	 	\textbf{Choise of gain parameter $\tau$ for CFW processing. } 
            \textbf{a.} Reduction of synthesized CFW error by time-average. Energy loss spectra of DMD structure in Fig. 2 of main text before CFW processing (blue curve), after CFW processing without time average (black curve) and after CFW processing with time average (red curve). The $\tau$ is \SI{3.5}{eV} for CFW processing.
            \textbf{b.} Energy loss spectra of DMD structure in Fig. 2 of main text processed by time-averaged CFW method with different gain $\tau$. 
            \textbf{c.}  Energy loss spectra of single silver nanoparticle in Fig. 3 of main text processed by time-averaged CFW method with different gain $\tau$.  
	    }
	\label{fig s_cfw processing}
\end{figure*}
%
In an EELS or CL experiment, the spectrum is acquired with a finite frequency range and discretized frequency points that can cause the synthesized CFW to form less steady decay than an ideal CFW, thus leading to oscillation errors in the CFW spectrum as shown in the black curve of Fig.~\ref{fig s_cfw processing}a. 
The solution to mitigate this interference is taking the time average of the CFW results~\cite{zeng2024synthesized}. 
As shown in the red curve of Fig.~\ref{fig s_cfw processing}a, the time-averaged CFW spectrum substantially reduces the finite-bandwidth induced signal oscillations, ensuring that the characteristic signals are sufficiently clear.

Another key parameter for the CFW technique is the gain factor $\tau$. Two limitations need to be noted: one is a lack of prior knowledge of actual optical loss $\gamma$ in the investigated system; the other one is that a too-large $\tau$ can also lead to the reappearance of the CFW oscillation errors (see Fig.~\ref{fig s_cfw processing}b and c) although the time-average method is applied. 
Therefore, we adopt a pragmatic selection of $\tau$. Specifically, We first commence with a smaller $\tau$ and progressively increase its value, evaluating and contrasting the CFW responses corresponding to different $\tau$ values. 
Upon this basis, we choose the CFW spectrum that demonstrates relatively substantial enhancement effects and whose oscillation errors do not interfere with the characteristic signals. 
As shown in Fig. 2b and Fig. 3b of the main text, we choose the CFW spectrum with $\tau$ = \SI{4}{\eV}  in Fig.~\ref{fig s_cfw processing}b and $\tau$ = \SI{0.9}{\eV}  in Fig.~\ref{fig s_cfw processing}c as the final results.

\section{Numerical simulations}
\label{sec:simulation}


For the DMD structure consisting of gold film sandwiched between two TiO$_2$ films in Fig. 2, the relative permittivity for $\mathrm{TiO}_\mathrm{2}$ from \SI{5}{\eV} to \SI{60}{\eV} were taken from the theoretical calculation in Ref.~\cite{vast2002local} and fitted using multi-coefficient models with the Lumerical software and the optical properties of gold in the same spectral regime were obtained from the tabulated data in Ref.~\cite{werner2009optical} with linear interpolation. 
For a gold film-coupled silver nanoantenna geometry in Fig. 4 studied in the visible regime, the optical properties of gold and silver were both obtained from the tabulated data in Ref.~\cite{johnson1972optical} with linear interpolation.

We performed the numerical simulations of the EELS spectrum in Fig. 2c of main text and CL spectrum in Fig. 4g using a finite-element electromagnetic solver (Radio Frequency Module of COMSOL Multiphysics).
Maxwell's equations were solved using the multifrontal massively parallel sparse direct solver (MUMPS).
The probing electron with velocity $v$ traveling in the $z$-direction was modeled as a line current of the form $\mathbf{j}(z,\omega) = - e \delta(\mathbf{R}-\mathbf{R}_0) \mathrm{e}^{\mathrm{i}\omega z/v} \mathbf{\hat{z}}$, where $\mathbf{R} = (x, y)$ is the transverse coordinate and $\mathbf{R}_0 = (x_0, y_0)$ is the electron position in the transverse plane.
The calculation for each frequency was simulated twice with and without the structure based on the same mesh to extract the total field $\mathbf{E}_{\mathrm{tot}}$ and incident field $\mathbf{E}_{\mathrm{inc}}$.
The scattered field is straightforwardly evaluated by their differences $\mathbf{E}_{\mathrm{sca}} = \mathbf{E}_{\mathrm{tot}}-\mathbf{E}_{\mathrm{inc}}$.
The electron energy loss probability $\Gamma(\omega)$ is calculated by the line integral of the induced field over the electron path, which is given as \cite{de2010optical} 
\begin{equation}
\begin{aligned}
\Gamma(\omega)
&=\frac{e}{\pi \hbar \omega }\int \diff{z} \Re{\mathrm{e}^{-\mathrm{i}\omega z/v} \mathbf{\hat{z}} \cdot  \mathbf{E}_{\mathrm{sca}}(z, \omega) } .
\label{eqsm:EELS}
\end{aligned}
\end{equation}

The CL probability $\Gamma_\mathrm{CL}(\omega)$ is calculated by the integration of the Poynting vector within a collection area $S$ corresponding to a collection angle between \SI{20}{\degree} and \SI{45}{\degree} of the Attolight CL-SEM:
\begin{equation}
\begin{aligned}
\Gamma_{{\rm CL}}(\omega)
&= \frac{1}{2\hbar\omega}\int_S \Re{ \mathbf{E}_{\mathrm{sca}}(\mathbf{r},\omega) \times \mathbf{H}_{\mathrm{sca}}^*(\mathbf{r},\omega) } \cdot  {\diff} \mathbf{S},
\label{eqsm:EELS}
\end{aligned}
\end{equation}
where $\mathbf{r} = (x, y, z)$ is the spatial coordinate and $\mathbf{H}_{\mathrm{sca}}$ is the scattered magnetic field. The resonant peaks in the CL spectra in Fig. 4 are analyzed using the eigenmode analysis of the film-coupled nanosphere under 2D simulations with a defined azimuthal integer number given the rotational symmetry of the system. The calculated eigenmodes are shown in Fig. 4h.

\section{Polarization-response representation of energy energy loss and cathdoluminescence}
\label{sec:EELS}
We adopt a $\Tmatrix$-matrix framework~\cite{zhang2023all,xie2024maximalquantuminteractionfree} to formulate general expressions of electron energy loss and cathodoluminescence. We treat free electrons interacting with an arbitrary photonic environment as an electromagnetic scattering problem~\cite{de2010optical,yang2018maximal,Miller2023}. In the polarization-response representation, the frequency-dependent extinction (\ie energy loss of free electrons) equals the work done by the incident field on the induced polarization field in the scatterer, $\Powerext(\omega)  = \left({\omega}/{2} \right) \mathrm{Im}\int_{\textit{V}}\Einc^{*}\cdot\mathbf{P}\,\mathrm{d}\mathit{V}$, which can formulate loss probability as
$\Gamma(\omega)
= \Powerext(\omega)/ \hbar\omega=
\frac{1}{ 2\hbar}\mathrm{Im}\int_{\textit{V}}\Einc^{*}\cdot\mathbf{P}\mathrm{d}\mathit{V}.
$
The polarization field can be expressed as the convolution of the incident field and one linear ``$\Tmatrix$ matrix" operator~\cite{zhang2023all,xie2024maximalquantuminteractionfree}: $\polfield(\mathbf{r}) = \int_V \Tmatrix(\mathbf{r},\mathbf{r^{\prime}}) \cdot \Einc( \mathbf{r^{\prime}} ) \mathrm{d} \mathbf{r^{\prime}}$ or $\mathbf{p} = \Tmatrix\einc$ (assuming identity vacuum permittivity $\epsilon_0=1$ that we use throughout) in vector-matrix notation.
Then, the loss probability can be expressed as
\begin{align}
\Gamma(\omega)
&=\frac{1}{2\hbar} \Im{\left[\einc^\dagger\Tmatrix(\omega)\einc\right]}=\frac{1}{2\hbar} \Im{\left\{ \Tr\left[\Tmatrix(\omega)\einc\einc^\dagger\right]\right\}}
\label{eq:T2ext}
\end{align}

Meanwhile, the frequency-dependent absorption $\Powerabs(\omega)$ is calculated by the work done by the total field $\Etot$ on the induced polarization by the scatterer: $\Powerabs(\omega) = 
\frac{\omega}{ 2}\Im{\int_{\textit{V}}\Etot^{*}\cdot\mathbf{P}\mathrm{d}\mathit{V}}$. %
For a scatter with material electric susceptibility $\chi(\Vx)$, the total field $\mathbf{E}_{\rm{tot}}(\Vx)$ can be described as: $\mathbf{P}_{\rm{tot}}(\Vx) = \chi(\Vx)\mathbf{E}_{\rm{tot}}(\Vx)$ or $\mathbf{p} = \chi\etot$ in vector-matrix notation, which can give rise to $\etot=\chi^{-1}\mathbf{p}=\chi^{-1}\Tmatrix\einc$. Then the frequency-dependent scattering can be described $\Powersca(\omega) = \Powerext(\omega) -\Powerabs(\omega)$ (Note that here we focus on the coherent cathodoluminescence instead of the incoherent one~\cite{roques2022framework,roques2023free}), which can formulate the CL probability as: 
\begin{equation}
\begin{aligned}
\Gamma_\mathrm{CL}(\omega)
&=\frac{\Powersca(\omega)}{\hbar\omega}=\frac{1}{2\hbar} \Im{\int_{\textit{V}}\Einc^{*}\cdot\mathbf{P}\mathrm{d}\mathit{V}}-\frac{1}{2\hbar}\Im{\int_{\textit{V}}\Etot^{*}\cdot\mathbf{P}\mathrm{d}\mathit{V}}\\
&=\frac{1}{2\hbar} \Im{(\einc^\dagger-\etot^\dagger)\Tmatrix(\omega)\einc}=\frac{1}{2\hbar} \Im{\Tr{[\mathbbm{I} -\Tmatrix^\dagger(\chi^{-1})^{\dagger}]\Tmatrix(\omega)\einc\einc^\dagger}}.
\label{eq:Scl}
\end{aligned}
\end{equation}
\section{Analytical energy loss and cathodoluminescence probability for canonical geometries}
\label{sec:analytical}

The optical permittivity of the half-space in the first analytical scenario (see Fig. 1b in main text) is described by the Drude model: $\varepsilon_{\rm{m}} = 1 - \omegap^2 / \omega(\omega + \iu \gamma_{\rm{m}})$, $\omegap = \SI{9.06}{\eV}$ for gold, $\gamma_{\rm{m}} = \SI{0.3}{\eV}$ for a high optical loss.
The optical permittivity of the nanosphere in the second analytical (see Fig. 1d in main text) scenario is described by the Drude model: $\varepsilon_{\rm{m}} = 1 - \omegap^2 /\omega(\omega + \iu \gamma_{\rm{m}})$, $\omegap = \SI{9.06}{\eV}$ for gold, $\gamma_{\rm{m}} = \SI{1.05}{\eV}$  for a high optical loss.

\subsection{Energy loss of electrons in half space}
A point electron with a constant speed of $v$ moves parallel to a half-space with a separation distance of $d$ along the $z$ axis in vacuum, as shown in Fig. 1b in the main text. 
The corresponding derivation of the energy loss has been well studied~\cite{lucas1971fast,echenique1975absorption,otto1967theory,garcia1985retardation,forstmann1991energy}, and the final expression can be written as:  
\begin{equation}
     \Gamma(\omega) =\frac{e^2\mathrm{L}}{2\pi^2\epsilon_0\hbar v^2}\int^\infty_0 \frac{\mathrm{d}k_x}{K^2} \Re \left\{ k_{y1} \mathrm{e}^{2 \mathrm{i} k_{y1}d} \left[ \left( \frac{k_xv}{k_{y1}c} \right)^2r_\mathrm{s}-r_\mathrm{p} \right] \right\},
\label{EELS-3D}
\end{equation}
where L is the length of electron trajectory, $K$ is the transverse wavenumber, $r_\mathrm{p}$ and $r_\mathrm{s}$ are reflection coefficients of the incident field with linearly polarized $\mathrm{p}-$ and $\mathrm{s}-$polarized plane waves, which are determined by 
\begin{align}
    K=\sqrt{ \frac{\omega^2} {v^2} + k_{x}^2 },
\label{smeq:tM-coefficient}
\end{align}
and
\begin{align}
   r_\mathrm{p}
&= \frac{\epsilon_2 k_{y1} - \epsilon_1 k_{y2}}{\epsilon_2 k_{y1} + \epsilon_1 k_{y2}}, 
\label{smeq:rp-coefficient}
\\
   r_\mathrm{s}
&= \frac{k_{y1} - k_{y2}}{k_{y1} + k_{y2}},
\label{smeq:rs-coefficient}
\end{align}

where $\epsilon_1$ and $\epsilon_2$ are the relative permittivity of vacuum and half-space, respectively, $k_{y1}=\sqrt{k_1 ^2 - K^2 }$, and $k_{y2}=\sqrt{k_2 ^2 - K^2 }$ based on the incident wavevector $\textbf{k}_1=(k_x,k_{y1},k_{z})$ and the transmitted wavevector $\textbf{k}_2=(k_x,k_{y2},k_{z})$.

\subsection{Cathodoluminescence of electrons near a sphere}
One electron with a constant speed of $v$ passes near a sphere with an impact parameter of $b$ as shown in Fig. 1d of the main text.
The corresponding CL emission probability can be derived based on the Mie theory~\cite{de2010optical}:
\begin{align}
    \Gamma_\mathrm{CL}(\omega) =\frac{e^2}{4\pi \epsilon_0 c \hbar\omega} \sum_{l=1}^{\infty} \sum_{m=-l}^{l} K_m^2 \left( \frac{\omega b}{v \gamma}\right) \left( C_{lm}^M \left|t_l^M\right|^2 + C_{lm}^E \left| t_l^E\right|^2 \right) ,
\label{EELS-sphere}
\end{align}
where $ K_m$ is the modified Bessel function of order $m$, $\gamma$ is the Lorentz factor, 
$C_{lm}^M$ and $C_{lm}^E$ represent the magnetic and electric coupling coefficients, and $t_l^M$ and $t_l^E$ are the magnetic and electric Mie scattering coefficients, which are determined by
\begin{align}
\gamma &=\frac{1}{\sqrt{1-(v/c)^2}}, 
\label{smeq:M-coefficient}
\\
   C_{lm}^M &=\frac{1}{l(l+1)} \left|2m N_{lm}\right|^2, 
\label{smeq:M-coefficient}
\\
   C_{lm}^E &=\frac{1}{l(l+1)} \left|\frac{c}{v\gamma} M_{lm}\right| ^2,
\label{smeq:E-coefficient}
\end{align}
with
\begin{align}
   N_{lm} &=\sqrt{\frac{(2l+1)}{\pi} \frac{(l-\abs{m})!}{(l+\abs{m})!} }\frac{(2\abs{m}-1)!!}{(v\gamma/c)^{\abs{m}}}C_{l-\abs{m}}^{(\abs{m}+1/2)}\left(\frac{c}{v} \right),
\label{smeq:N-coefficient1}
\\
   M_{lm} &= N_{lm+1}\sqrt{(l+m+1)(l-m)}+N_{lm-1}\sqrt{(l-m+1)(l+m)},
\label{smeq:M-coefficient1}
\end{align}
and
\begin{align}
    t_l^M &= \frac{-j_{l}(x_1) x_2 j_l^{\prime}(x_2) + x_1 j_l^{\prime}(x_1) j_l(x_2) }{h_l^{(+)}(x_1) x_2 j_l^{\prime}(x_2) - x_1 \left[ h_l^{(+)}(x_1) \right]^{\prime} j_l(x_2) },
\label{smeq:tM-coefficient}
\\
    t_l^E &= \frac{ -\epsilon_1 j_l(x_1) \left[ x_2 j_l(x_2)\right]^{\prime} + \epsilon_2 \left[ x_1 j_l(x_1) \right]^{\prime} j_l(x_2) }{\epsilon_1 h_l^{(+)}(x_1)  \left[ x_2 j_l(x_2)\right]^{\prime}  - \epsilon_2 \left[x_1 h_l^{(+)}(x_1) \right]^{\prime} j_l(x_2) },
\label{smeq:tE-coefficient}
\end{align}
where $\epsilon_1$ and $\epsilon_2$ are the relative permittivities of background and the sphere, $l$ represent multipolar order of the excitation associated with the order of Legendre polynomials ($l =1 $ for dipole excitation, $l =2 $ for quadrupole excitation), $m$ denotes the azimuthal index of the excitation, $C_{l-\abs{m}}^{(\abs{m}+1/2)}$ is Gegenbauer polynomial, primed functions stand for their derivatives, $x_1 = \omega a \sqrt{\epsilon_1} /c$, $x_2 = \omega a \sqrt{\epsilon_2} /c$, and $j_l$ and $ h_l^{(+)} = \mathrm{i} j_l - y_l$ correspond to spherical Bessel and Hankel functions, respectively.

\section{Complex frequency waves for electron energy gain spectroscopy}
\label{sec:EEGS}

\begin{figure*}[!htbp]
	\centering
    \includegraphics[width=1\linewidth]{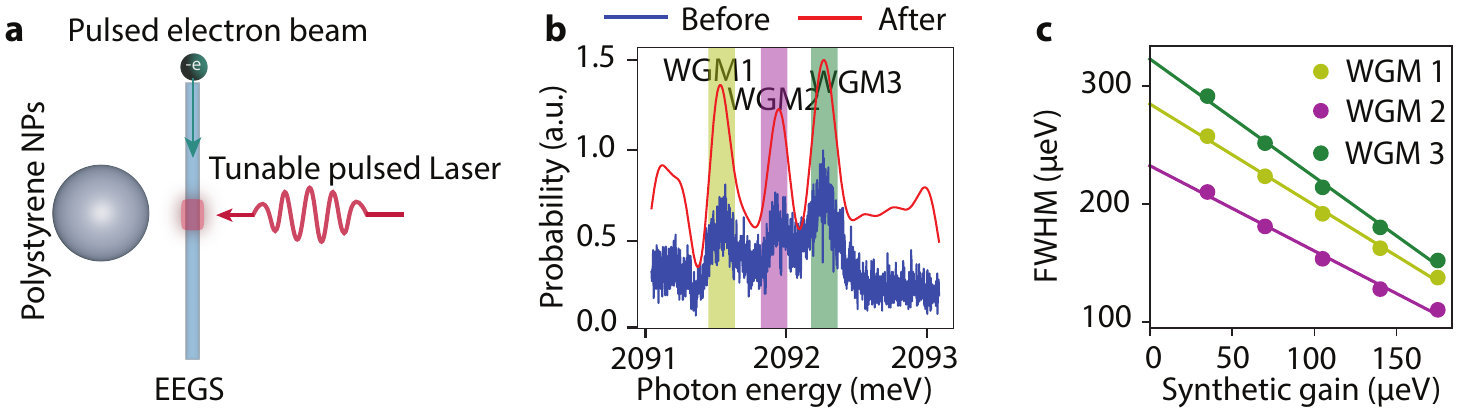}
	 \caption{%
            \textbf{Characterizing neighboring high--quality-factor modes of a single polystyrene nanoparticle in electron energy gain spectroscopy (original measurement data taken from Ref.~\cite{auad2023muev}). } 
            \textbf{a.} Schematic of EEGS measurement. The pulsed electron beam interacts with one polystyrene nanosphere illuminated by a tunable pulsed laser. 
            \textbf{b.} EEGS spectra before and after CFW processing, constructed from EEGS signals as a function of the laser wavelength with a step size of 0.4 pm. The gain parameter $\tau$ is chosen as $\SI{175}{\mu\eV}$.
            \textbf{c.} Full-width half maximum of the three modes (dots) and their linear fitting (lines) as a function of the synthetic gain $\tau$ under CFW processing.  
            }
	    
	\label{fig_main:S1}
\end{figure*}
%
Aside from EELS and CL analyzed in the main text, here we show the potential utility of the CFW technique in electron energy gain spectroscopy (EEGS)~\cite{de2008electron}. Similar to CL, EEGS combines ultra-high spectral resolution (same as that of optical spectroscopy) with the sub-nanometer spatial resolution of electron microscopy. Moreover, EEGS enjoys a higher signal-to-noise ratio than CL because of the external illumination. Here, we apply the CFW processing to the data reported by \citet{auad2023muev}. 

200-keV electrons pass by single polystyrene microspheres with the illumination of a tunable pulsed laser as shown in Fig.~\ref{fig_main:S1}a. 
The original EEGS spectrum (blue curve of Fig.~\ref{fig_main:S1}b; obtained from Ref.~\cite{auad2023muev}) displays three major peaks corresponding to three whispering-gallery modes (WGMs).
Their resonance features, \eg the quality factor, are usually extracted by a summation of Lorentizian fitting in practice.

In the following, we show that it is possible to extract the quality factors of these neighboring resonances using CFW under various virtual gain parameters $\tau$.
Because the full-width at half-maximum (FWHM) of a Lorentzian peak equals twice the imaginary part of the complex resonant frequency, the FWHM of each resonance should decreases linearly as the synthetic gain $\tau$ increases.
Let us first look at a single-run spectrum at fixed $\tau$.
After CFW processing, the spectral features of the three WGMs are enhanced (red curve of Fig.~\ref{fig_main:S1}b). 
Therefore, one can monitor the linewidths of these peaks under various virtual gain $\tau$ (Fig.~\ref{fig_main:S1}c).
The linewidths of all three modes exhibit a linear trend as expected, and they are fitted with linear regression, where the crossing of the linear fitting with the $y$ axis in Fig.~\ref{fig_main:S1}c provides an estimate of the quality factors of the passive cavity as measured under $\tau=0$. 
In Fig.~\ref{fig_main:S1}c, the quality factors of the modes WGM1, WGM2, and WGM3 are extracted from our CFW approach as $7293\pm 66$, $8946 \pm 129$, and $6432 \pm 47$, respectively.

\section{Spontaneous multi-photon events in free-electron--photon quantum interaction}
\label{sec:Poisson}
Free electron's spontaneous emission of multiple photons obeying Poisson statistics
\begin{equation}
    \Gamma(\omega) = \sum_{\textit{i}=1}^{\infty}{\mathrm{e}^{-\lambda}\frac{\lambda^n}{n!}f_n(\omega)},
\label{EELS-3D}
\end{equation}
where $\Gamma(\omega)$ is the total loss probability accounting all orders of photon events, $\lambda=\abs{\gqu}^2$ is the Poisson parameter ($\gqu$ is the quantum interaction strength), $n$ is the photon number. $f_n(\omega)$ is the spectral density of the $n^{th}$ loss peak, which can be obtained with a recursive formula~\cite{adiv2023observation}:
\begin{equation}
    f_n(\omega)=\int f_{n-1}(\omega-\omega^{\prime})\Gamma_0(\omega^{\prime}) \mathrm{d} \omega^{\prime}; n=1,2,3,...
\label{eq:int-inc-field}
\end{equation}
where $\Gamma_0(\omega)$ is the electron energy loss modelled with a Lorentzian spectral shape (\eg for a single bosonic mode the electron couples to) corresponding to the one-photon event only. $f_0(\omega)$ relates to the initial energy distribution of electrons (zero-loss peak) and is modeled by a Gaussian distribution.
%


%